\documentclass[onecolumn]{revtex4}

\usepackage{graphicx}
\usepackage{dcolumn}
\usepackage{color}
\usepackage{bm}
\usepackage{amsmath}
\usepackage{amsfonts}
\usepackage{amssymb}

\input epsf

\begin{document}

\title{An Alternative f (R, T ) Gravity Theory and the  Dark Energy Problem}

\author{Subenoy Chakraborty\footnote{schakraborty@math.jdvu.ac.in}}

\affiliation{Department of Mathematics, Jadavpur University, Kolkata-700 032, India.}{}
\begin{abstract}
Recently, a generalized gravity theory was proposed by Harko etal where the Lagrangian density is an arbitrary function of the Ricci scalar R and the trace of the stress-energy tensor T, known as F(R,T) gravity. In their derivation of the field equations, they have not considered conservation of the stress-energy tensor. In the present work, we have shown that a part of the arbitrary function f(R,T) can be determined if we take into account of the conservation of stress-energy tensor, although the form of the field equations remain similar. For homogeneous and isotropic model of the universe the field equations are solved and corresponding cosmological aspects has been discussed. Finally, we have studied the energy conditions in this modified gravity theory both generally and a particular case of perfect fluid with constant equation of state.\\

\end{abstract}


\maketitle

\section{Introduction}
Recent observational predictions [1] that our universe is going through a phase of accelerated expansion put new avenues in modern cosmology. A class of people are making attempts to accomodate this observational fact by choosing some exotic matter(known as dark energy) in the framework of general relativity.There are several choices for this exotic matter namely a) the quintessence scalar field models[2], the phantom field[3], K-essence [4], tachyon field [5], quintom [6] etc., b) the dark energy models including Chaplygin gas[7] and so on. On the other hand, there are attempts to modify the gravity theory itself to accomodate the present accelerated phase. A natural generalization is to choose a more general action in which the standard Einstein-Hilbert action  is replaced by an arbitrary function of the Ricci scalar R [8] (i.e, f(R)) and is known as f(R) -gravity. This modified theory may explain this late time cosmic acceleration [9]. These $f(R)$ models can satisfy local tests and unify inflation with dark energy[10]. Also it is possible to explain the galactic dynamics of massive test particles in this modified gravity theory without any dark matter [11-15]. For detailed review of f(R)- gravity one may refer to [8, 16].\\

 Recently, a further generalization of $f(R)$ - gravity theory has been done by Harko etal [17]. They choose the Lagrangian density as arbitrary function $f(R,T)$ where as usual R is the Ricci scalar and T is the trace of the energy -momentum tensor. The justification of choosing T as an argument for the Lagrangian is from exotic imperfect fluids or quantum effects (conformal anomaly). They have argued that due to the coupling of the matter and geometry, this gravity model depends on a source term, which is nothing but the variation of the matter stress-energy tensor. As a result, the motion of test particles is not along geodesic path due to the presence of an extra force perpendicular to the four velocity. The cosmic acceleration in this modified $f(R,T)$ theory results not only from geometrical contribution but also from the matter content. Subsequently, Houndjo [18] has chosen $f(R,T)$ as $f_{1}(R) + f_{2}(T)$ and discussed transition of matter dominated era to an accelerated phase. Very recently, Sharif etal [19] have studied thermodynamics in this $f(R,T)$ theory and Azizi [20] have examined the possibility of wormhole geometry in $f(R,T)$ gravity.\\

In the present article, we have formulated the $f(R,T)$ gravity theory in an unorthodox manner. Though the action is a coupling of geometry and matter, but still we restrict ourselves to the special cases where test particles move in a geodesics. As a result, the Lagrangian has some restricted form, keeping the field equations same. The alternative derivation of $f(R,T)$ gravity and some specific choice for $f(R,T)$ has been presented in section II. Also admissibility of some known matter fields has been examined in this section. Section III deals with cosmological solutions for homogeneous and isotropic model of the universe with some physical interpretations. Energy conditions in this modified gravity theory has been examined both in a general way as well as for perfect fluid in section IV. Finally, at the end there is a brief summary of the entire work in section V.
\section{$f(R,T)$ gravity theory: A modification}
In this gravity theory [17], the gravitational Lagrangian density is given by an arbitrary function $f(R,T)$ of two variables: One is the Ricci scalar R and the other is the trace of the energy-momentum tensor $T(=T_{\mu \nu}g^{\mu \nu})$. So the complete action of this theory is written as [17]

\begin{equation}
 A = \frac{1}{16\Pi}\int f(R,T)\sqrt{- g} d^{4}x + \int L_{m} \sqrt{-g} d^{4}x
\end{equation}

where the stress-energy tensor of the matter $(T_{\mu\nu})$ can be obtained from the matter Lagrangian density $L_{m}$ as [21]

\begin{equation}
T_{\mu\nu} =-\frac{2}{\sqrt{-g}} \frac{\delta(\sqrt{-g}L_{m})}{\delta g^{\mu\nu}}
\end{equation}

This can be simplified further assuming $L_{m}$ depends only on $g_{\mu\nu}$ but not on its derivatives as

\begin{equation}
T_{\mu\nu} = g_{\mu\nu} L_{m}- 2 \frac{\partial L_{m}}{\partial g^{\mu\nu}}
\end{equation}

Using the standard text book result namely

\begin{equation}
\delta R = R_{\mu \nu} \delta g^{\mu \nu} + g^{\mu \nu}
\Box \delta g^{\mu \nu} - \nabla_{\mu}\nabla_{\nu} \delta g^{\mu \nu}
\end{equation}

and the shortcut notations:

\begin{equation}
f_{R} = \frac{\partial f(R,T)}{\partial R},  f_{T} = \frac{\partial f(R,T)}{\partial T}
\end{equation}

the variation of the above action can be written as

\begin{equation}
\delta A = \frac{1}{16\Pi} \int [ f_{R} (R_{\mu \nu} \delta g^{\mu \nu} + g_{\mu \nu} \Box \delta g^{\mu \nu} - \nabla _{\mu} \nabla _{\nu} \delta g^{\mu \nu}) + {f_{T}\frac{\delta (g^{\alpha \beta}T_{\alpha \beta})}{\delta g^{\mu \nu}} - \frac{1}{2}g_{\mu \nu}f(R,T)}\delta g^{\mu \nu}
+ \frac{16\Pi}{\sqrt{-g}} \frac{\delta (\sqrt{-g} L_{m})}{\delta g^{\mu \nu}}] \sqrt{-g} d^{4} x .
\end{equation}

Now performing by parts integration to the second and third terms in the r.h.s. of equation (6), one obtains the field equations in $f(R,T)$ gravity theory as [17],

\begin{equation}
f_{R}R_{\mu \nu} - \frac{1}{2} f(R,T) g_{\mu \nu} + ( g_{\mu \nu}\Box - \nabla _{\mu} \nabla _{\nu} )f_{R} = 8 \Pi T_{\mu \nu} -
f_{T}( T_{\mu \nu} + \Theta _{\mu \nu})
\end{equation}

with

\begin{equation}
\Theta _{\mu \nu} = g^{\alpha \beta} \frac{\delta T_{\alpha \beta}}{\delta g^{\mu \nu}} = -2 T_{\mu \nu} + g_{\mu \nu} L_{m} - 2 g^{\alpha \beta}\frac{\partial ^{2} L_{m}}{\partial g^{\mu \nu} \partial g^{\alpha \beta}}
\end{equation}

It is to be noted that if $f(R,T) = f(R)$ then we get back to the field equations for f(R) gravity.\\

        Now we can proceed further, with the field equations (7) in the following three cases:\\
        a)$f(R,T) = R + h(T)$\\

        b)$f(R,T) = R .h(T)$\\

        c)$f(R,T)$ is arbitrary\\

 $\bullet$  Case-(a) :     $f(R,T) = R + h(T)$\\
   For this choice of $f(R,T)$ the field equations (7) now simplify to

\begin{equation}
G_{\mu \nu} = 8 \Pi T_{\mu \nu} - h\prime(T) ( T_{\mu \nu} + \Theta _{\mu \nu}) + \frac{1}{2} h(T) g_{\mu \nu}
\end{equation}

Now taking  divergence of both sides of the above field equations(9) and assuming conservation of energy -momentum tensor (i.e, $\nabla ^{\mu} T_{\mu \nu} = 0$) we obtain

\begin{equation}
(T_{\mu \nu} + \Theta _{\mu \nu})\nabla ^{\mu} h\prime(T) + h\prime(T)\nabla ^{\mu} \Theta _{\mu \nu} +  \frac{1}{2} g_{\mu \nu} \nabla ^{\mu} h(T) = 0
\end{equation}

   This shows that the form of h(T) is not arbitrary, it depends on the choice of the matter field.
   We consider now some known matter fields as examples:\\
   Example-I:  Electromagnetic Field.\\

 The matter Lagrangian has the form

 \begin{equation}
 L_{m} = -\frac{1}{16 \Pi } F_{\alpha \beta} F_{\gamma \sigma} g^{\alpha \gamma} g^{\beta \sigma}
 \end{equation}

 with $F_{\mu \nu}$, the electromagnetic field tensor. So from equation (8) we have

 \begin{equation}
 \Theta_{\mu \nu} = -T_{\mu \nu}.
 \end{equation}

 As a result equation (10) simplifies to
                   $\frac{\partial h(T)}{\partial x^{\mu}} = 0$
  i.e, $h(T)$ turns out to be a constant. Thus for this choice of $f(R,T)$ electromagnetic field is not possible.\\

  Example-II:  Perfect fluid.\\

In case of perfect fluid, the stress-energy tensor has the usual form

 \begin{equation}
 T_{\mu \nu} = (\rho + p) u_{\mu} u_{\nu} - p g_{\mu \nu}
 \end{equation}

 and the matter Lagrangian can be taken as $L_{m} = -p$. Here $\rho$ and p are the usual energy density and thermodynamic pressure and the four velocity $u^{\mu}$ satisfies
 i) $u_{\mu} u^{\mu} =1$ and ii) $u^{\mu} \nabla_{\nu} u_{\mu} = 0$.
 In this case $\Theta_{\mu \nu}$ has the explicit form

 \begin{equation}
 \Theta_{\mu \nu} = -2 T_{\mu \nu} - p g_{\mu \nu}.
 \end{equation}

 Now substituting equation (14) for $\Theta_{\mu \nu}$ into equation (10) we obtain

 \begin{equation}
 (T_{\mu \nu} + p g_{\mu \nu}) \nabla^{\mu} h\prime(T) + h\prime(T) g_{\mu \nu} \nabla^{\mu} p + \frac{1}{2} g_{\mu \nu} \nabla^{\mu} h(T) = 0.
 \end{equation}

 Further, if the perfect fluid has barotropic equation of state, i.e, $p = \omega \rho$, $\omega$, a constant then for homogeneous and isotropic flat FRW model $h(T)$ has an explicit form as $(\omega \neq -1,\pm \frac{1}{3})$

 \begin{equation}
 h(T) = h_{0} T^{\alpha}
 \end{equation}

 where $\alpha = \frac{1+ 3 \omega }{2(1 + \omega)}$, $h_{o}$ is an integration constant.\\
$\bullet$ Case (b):

For the choice $f(R,T) = R h(T)$, the field equations become

 \begin{equation}
 G_{\mu \nu} = \frac{8\Pi T_{\mu \nu}}{h(T)} - \frac{R h\prime(T)}{h(T)} (T_{\mu \nu} + \Theta_{\mu \nu}).
 \end{equation}

 After taking divergence of both sides of equation (17) and considering energy conservation relation,one obtains the differential equation
 for $h(T)$ as

 \begin{equation}
 8\Pi T_{\mu \nu} \nabla^{\mu}(\frac{1}{h(T)} - \nabla^{\mu}(\frac{h\prime(T)R}{h(T)})(T_{\mu \nu} + \Theta_{\mu \nu}) - \frac{h\prime(T)R}{h(T)} \nabla^{\mu} \Theta_{\mu \nu} = 0,
 \end{equation}

 where Ricci scalar R is related to T and h by the relation

 \begin{equation}
 R = 8 \Pi T/ [h\prime(T)(\theta + T) - h(T)].
 \end{equation}

 Note that here also for electromagnetic field h(T) is restricted by the relation $\frac{\partial h(T)}{\partial x^{\mu}} = 0$ i.e,  $h(T)$ is a constant.

 $\bullet$ Case (c):

Here $f(R,T)$ is totally arbitrary except the choices in the previous two cases. We start with the geometric identity namely [22]

$$(\Box \nabla _{\nu} - \nabla _{\nu} \Box )f(R,T) = R_{\mu \nu} \nabla^{\mu} f(R,T)$$
 i.e.

 \begin{equation}
 \nabla ^{\mu}(\nabla _{\mu} \nabla _{\nu} - g_{\mu \nu} \Box ) f(R,T) = R_{\mu \nu} \nabla^{\mu} f(R,T).
 \end{equation}

 Now taking covariant divergence of equation (7) and using this identity we have from conservation of matter field

 \begin{equation}
 (T_{\mu \nu} + \Theta _{\mu \nu}) \nabla^{\mu} f_{T} + f_{T} \nabla ^{\mu} \Theta_{\mu \nu} = 0.
 \end{equation}

 Note that equation (21) is not identical in form to that of equation (10), there is one extra term in equation (10). Now proceeding as before we
 obtain the following results:\\

 $\bullet$\textbf{I}. In case of electromagnetic field equation(21) is identically satisfied and hence $f_{T}$ is an arbitrary function of R and T.
 Thus a general form of $f(R,T)$ can be written as

 \begin{equation}
 f(R,T) = A_{0}(R,T) + A_{1}(R)
 \end{equation}

 where $A_{0}$ and $A_{1}$ are arbitrary functions of arguments.\\

 $\bullet$\textbf{II}. Similarly, for perfect fluid the form of $f(R,T)$ turns out to be

 \begin{equation}
 f(R,T) = A(R) + B(T)
 \end{equation}

 where $A(R)$ is an arbitrary function of R (except $A(R) = R$) and $B(T)$ has the form

 \begin{equation}
 B(T) = B_{0}\int exp[-\int\frac{dp}{\rho + p}] dT
 \end{equation}

 with $B_{0}$, an integration constant. If the fluid is in barotropic nature with constant equation of state then we have

 \begin{equation}
 B(T) = B_{0} T^{\alpha}, \alpha = \frac{1}{1+ \omega} (\omega \neq -1).
 \end{equation}

 Thus the choice of the function $f(R,T)$ depends to a great extend on the matter field taken into account.

\section{Cosmological Solutions and Consequences}

We now try to find cosmological solutions for the first choice of $f(R,T)$ for perfect fluid in the background of flat FRW model. The Einstein field equations are

\begin{equation}
3 H^{2} = \rho + h_{0}(1-3 \omega )^{\alpha - 1} \rho ^{\alpha}
\end{equation}

and

\begin{equation}
2 \dot{H}+ 3 H^{2} = - p + \frac{1}{2} h_{0}(1- 3 \omega )^{\alpha} \rho^{\alpha}
\end{equation}

where we have used the solution (16) for $h(T)$ .

In Einstein gravity, the above field equations correspond to a non-interacting two-fluid system of which one is the usual perfect fluid (Fluid -1) that we have considered  in $f(R,T)$ -gravity theory while the other fluid system (Fluid-2) is also a perfect fluid having energy density and pressure

\begin{equation}
\rho_{d} = h_{0} ( 1-3 \omega )^{\alpha -1} \rho^{\alpha} , p_{d} = -\frac{1}{2}h_{0} (1-3 \omega )^{\alpha} \rho^{\alpha}
\end{equation}

The equation of state  of the additional fluid (i.e. Fluid-2) is

\begin{equation}
\omega_{d} = \frac{p_{d}}{\rho_{d}} = -\frac{1-3 \omega}{2}
\end{equation}

with
$$\rho_{d} + p_{d} = \frac{h_{0}}{2}(1+3 \omega )(1-3 \omega )^{\alpha-1} \rho^{\alpha}$$
and

\begin{equation}
\rho_{d} +3 p_{d} = \frac{h_{0}}{2}(9 \omega -1 )(1-3 \omega )^{\alpha-1} \rho ^{\alpha},
\end{equation}

The nature of the two non-interacting fluids in different stages of the evolution of the universe are shown in table I:\\
\begin{table}
\begin{center}

\caption{\bf  Evolution of the Universe and the nature of the 2-fluids}
\centering

\begin{tabular}{|c|c|c|}
\hline
{\bf Evolution of the universe}& {\bf Fluid-1} & {\bf Fluid-2} \\[0.3ex]
\hline

Ultra relativistic era & $\omega = 1$(Stiff fluid) & $\omega_{d} = 1$ (Stiff fluid)\\
Early Universe before radiation & Fluid in pre-radiation era($\frac{1}{3} < \omega \leq 1$) & Normal fluid with +ve pressure($0 <\omega_{d}\leq1$)\\
Radiation era & $\omega = \frac{1}{3}$ & $\omega_{d} =0$ :dust\\
After radiation era & $\frac{1}{9} < \omega < \frac{1}{3}$ & - ve pressure but non-exotic in nature i.e. satisfies SEC\\
Till dust era & $0 \leq \omega < \frac{1}{9}$ & Exotic fluid with $-\frac{1}{2}\leq \omega_{d} \leq -\frac{1}{3}$\\
Before quintessence era & $-\frac{1}{3} \leq \omega < 0$ & Exotic fluid upto phantom barrier\\
Quintessence era & $-1 < \omega < -\frac{1}{3}$ & Phantom fluid\\
Phantom era & $\omega < -1$ & Ultra phantom fluid\\
\hline

\end{tabular}
\end{center}
\end{table}

Thus, although both fluids start simultaneously at the ultra-relativistic equation of state (stiff fluid) but fluid-2 advances more  rapidly so that it reaches the quintessence equation of state when the actual fluid (i.e, Fluid-1) has still positive pressure and finally fluid-2 reaches the phantom era when the fluid-1 is in quintessence era. Further, it is to be noted that although both the fluid components have constant equation of state but the effective one fluid system has always variable equation of state.\\
As for the normal fluid (i.e.Fluid-1) we have $p = \omega \rho$ so from the conservation of energy - momentum tensor, i.e, $$\dot{\rho} + 3H( \rho + p) = 0$$

we have on integration,
\begin{equation}
\rho = \rho_{0} a^{-3(1 + \omega )}
\end{equation}

where $\rho_{0}$ is an integration constant. Now substituting  this value of $\rho$ into the Friedmann equation (26) we obtain an integral equation for the scale factor 'a' as

\begin{equation}
\pm (t-t_{0}) = \int \frac{a^{\frac{1 + 3 \omega}{2}}da}{\surd [d_{1} + d_{2} a^{\frac{3(1- \omega)}{2}}]}
\end{equation}

with $t_{0}$ an integration constant and $d_{1} = \frac{8 \Pi \rho_{0}}{3},  d_{2} = \frac{h_{0} \rho_{0}^{\alpha}(1- 3 \omega)^{(\alpha - 1)}}{3}$.
In the following we have explicit solution for 'a' with $\omega = 0, \pm 1$ as

\begin{equation}
a^{\frac{3}{2}} = \frac{9d_{2}}{16}(t-t_{0})^{2} - \frac{d_{1}}{d_{2}} ,~~~ \omega = 0
\end{equation}

\begin{equation}
a^{3} = a_{0} (t - t_{0}) ,~~~  \omega = 1
\end{equation}

and

\begin{equation}
 a^{-\frac{3}{2}} = \frac{9 d_{1}}{16} (t - t_{0})^{2} - \frac{d_{2}}{d_{1}} ,~~~ \omega =-1.
\end{equation}

Note that on the phantom barrier ( i.e.$ \omega = -1$) we have a big rip singularity at finite time $t = t_{0} + \frac{4 \sqrt{d_{2}}}{3 d_{1}}$. The other two  solutions are the usual expanding solutions starting from the big-bang singularity at finite past.

\section{Energy Conditions in $f(R,T)$ Gravity}

The fundamental features to the singularity theorems  as well as to those related to classical black hole thermodynamics [23] are nothing but the energy conditions which are consequences of the Raychaudhuri equation for expansion, namely,

\begin{equation}
\frac{d \theta }{d \tau } = -\frac{1}{2} \theta^{2} - \sigma_{\mu \nu} \sigma^{\mu \nu} + \omega_{\mu \nu} \omega^{\mu \nu} - R_{\mu \nu} \kappa^{\mu} \kappa^{\nu}
\end{equation}

Here $\theta$, $\sigma_{\mu \nu}$ and $\omega_{\mu \nu}$ are respectively the expansion, shear, and rotation associated to the congruence defined by the null vector field $\kappa^{\mu}$ and $R_{\mu \nu}$ is the usual Ricci tensor. Though the Raychaudhuri equation is not related to any gravity theory (it is purely a geometric statement) but it has some special reference to Einstein gravity. As the attractive character of gravity is reflected through the positivity condition, i.e, $R_{\mu \nu} \kappa^{\mu} \kappa^{\nu} \geq 0$ (which implies that the geodesic congruences focus within a finite value of the parameter labeling points on the geodesics [24] ), so in Einstein gravity the above condition becomes $T_{\mu \nu} \kappa^{\mu} \kappa^{\nu} \geq 0$, which is the null energy condition (NEC). The weak energy condition (WEC), i.e, $T_{\mu \nu} v^{\mu} v^{\nu} \geq 0$ ($v^{\mu}$, a time-like vector) assumes the positivity of the local energy density and by continuity, $WEC \Rightarrow NEC$. Similarly, we have two other energy conditions namely the strong energy condition (SEC): $(R_{\mu \nu} - \frac{1}{2} R g_{\mu \nu} ) v^{\mu} v^{\nu} \geq 0$ which by continuity implies NEC but not the WEC in general and the dominant energy condition (DEC): $T_{\mu \nu}v^{\mu} v^{\nu}\geq 0$ and $T_{\mu \nu} v^{\nu}$ is not space-like imply locally measured energy density to be always positive and the energy flux is time-like or null. Also $DEC \Rightarrow WEC $(and hence the NEC) but not necessarily the SEC ( For details of energy conditions see [25]). For perfect fluid the above energy conditions have the explicit form:\\

\begin{eqnarray}
\left. \begin{array}{c}
NEC :~~~  \rho + p \geq 0\\
WEC :~~~  \rho \geq 0 ,~~~~ \rho + p \geq 0\\
SEC :~~~   ( \rho  + 3 p) \geq 0 , ~~~~\rho + p\geq 0\\
DEC :~~~ \rho \geq 0   and  ~~~~ \rho \pm p \geq 0
\end{array} \right \}
\end{eqnarray}

\begin{figure}

\includegraphics[height=3in, width=3in]{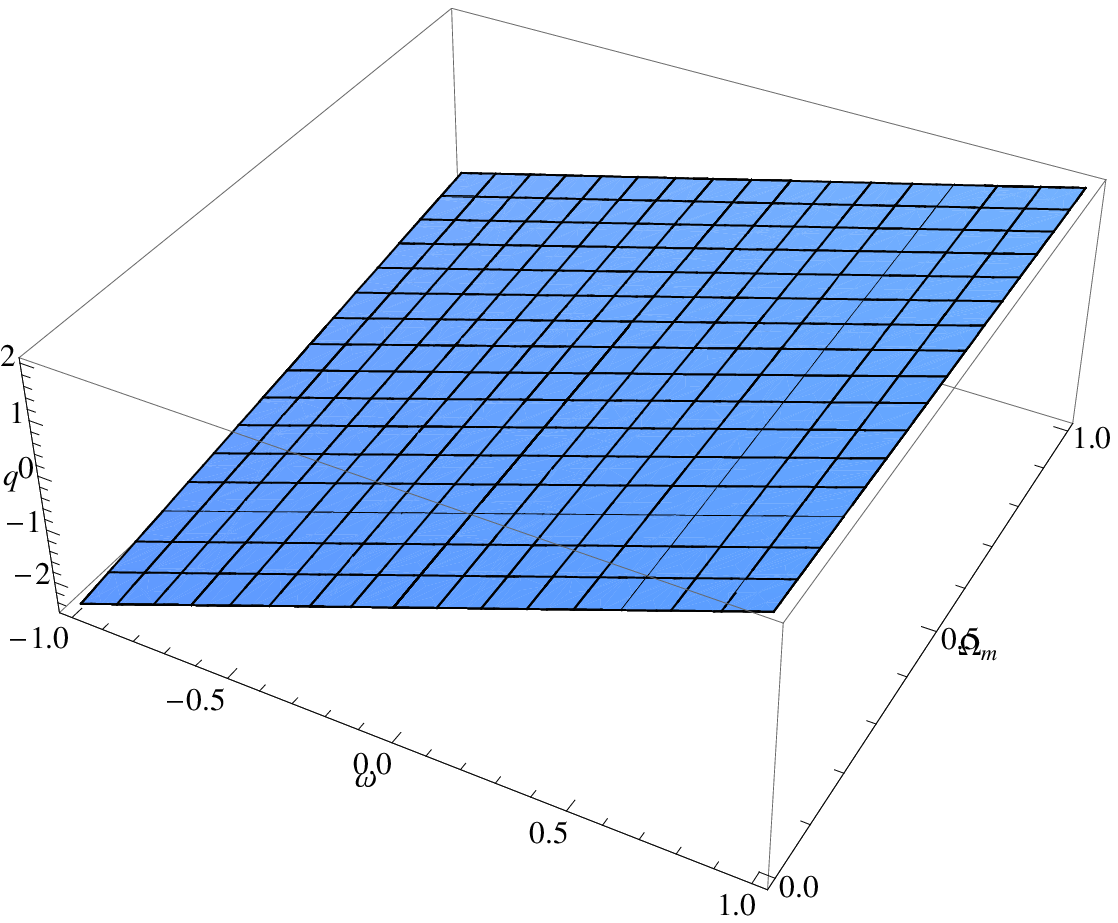}~~~

\vspace{1mm}Figure 1: Shows the graphical representation of the deceleration parameter $q$ for the variation of $\omega$ and $\Omega _{m}$.\hspace{1cm}
 \vspace{1mm}

\end{figure}

\begin{figure}

\includegraphics[height=3in, width=3in]{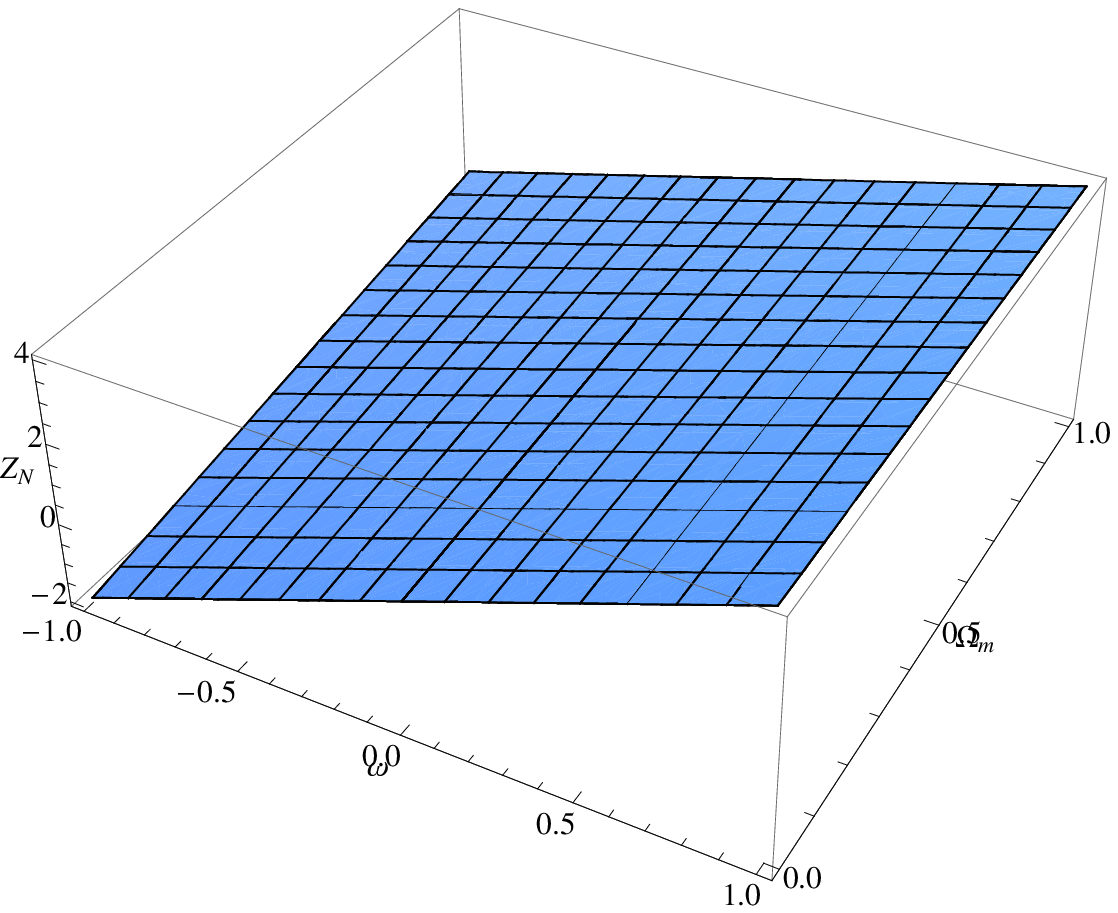}~~~

\vspace{1mm}Figure 2: Represents the variation of $Z_{N}=(1-\omega)\Omega _{m}+(1+3\omega)$ against $\omega$ and $\Omega _{m}$. The range $Z_{N}\geq 0$ represents validity of $NEC$.\hspace{1cm}
 \vspace{1mm}

\end{figure}
\begin{figure}

\includegraphics[height=3in, width=3in]{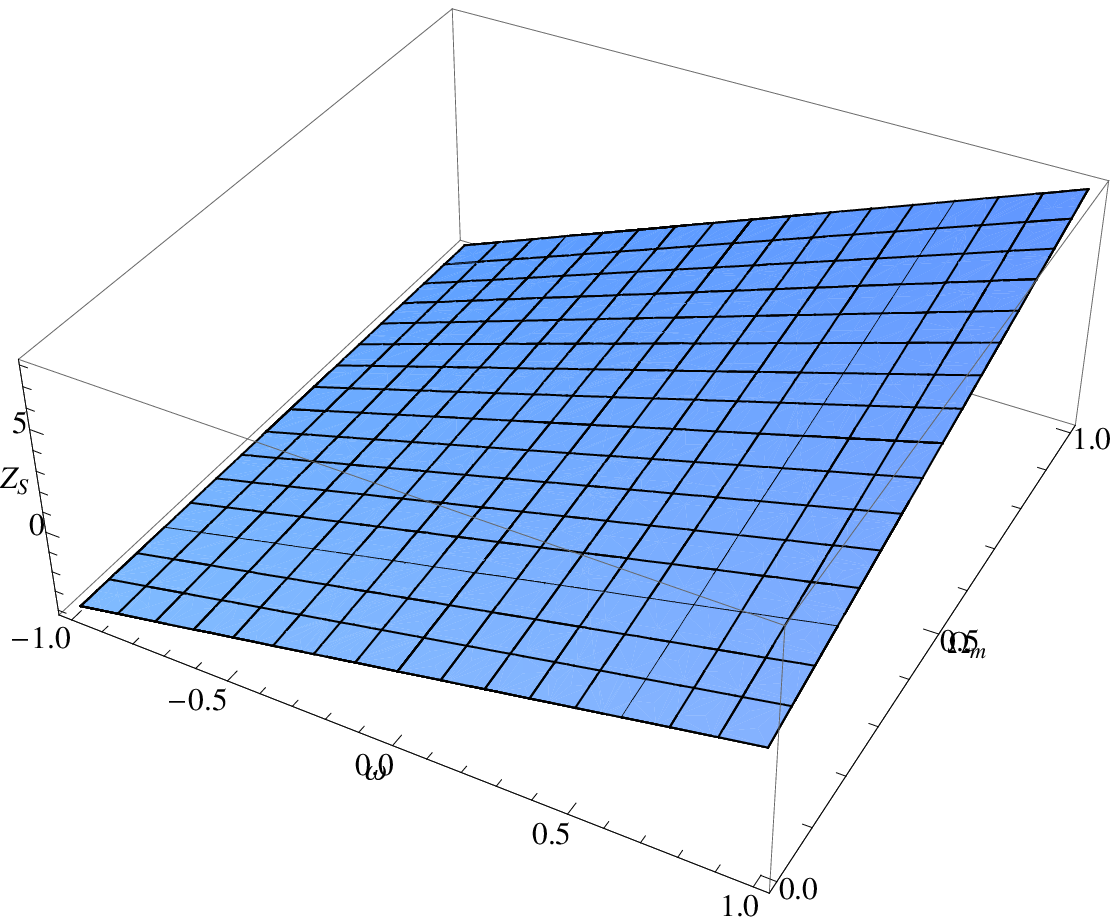}~~~

\vspace{1mm}Figure 3: Plots the variation of $Z_{S}=3(1+\omega)\Omega _{m}-(1-3\omega)$ against $\omega$ and $\Omega _{m}$. The range $Z_{S}\geq 0$ stands for the validity of $SEC$.\hspace{1cm}
 \vspace{1mm}

\end{figure}

But difficulty arises in other gravity theories, particularly where $R_{\mu \nu}$ may not be evaluated using the corresponding field equations. However, in a gravity theory if the Lagrangian density still have an Einstein-Hilbert term then it is possible to determine $R_{\mu \nu} \kappa^{\mu} \kappa^{\nu}$.\\

~~~~~~~~In the present $f(R,T)$ gravity theory, for the first two choices of $f(R,T)$ (i.e. $f(R,T) = R + h(T)~ or ~R h(T)$) the field equations (9) or (17) can be written as,

\begin{equation}
G_{\mu \nu} = T_{\mu \nu}^{eff}
\end{equation}

and the energy conditions read as, \\

\begin{eqnarray}
\left. \begin{array}{c}
NEC: ~~~  \rho_{eff} + p_{eff} \geq 0\\
WEC: ~~~  \rho_{eff} \geq 0 , ~~~\rho_{eff} + p_{eff} \geq 0\\
SEC: ~~~  \rho_{eff} + 3 p_{eff} \geq 0\\
and\\
~~~\rho_{eff} + p_{eff} \geq 0\\
DEC: ~~~  \rho_{eff} \geq 0 and~~~ \rho_{eff} \pm p_{eff} \geq 0\\
\end{array}\right \}
\end{eqnarray}\\

However, explicitly if we consider the perfect fluid case of the previous section then from the field equations (26) and (27) we have

$$\rho_{eff} = \rho + h_{0} ( 1 - 3 \omega )^{(\alpha - 1)} \rho^{\alpha}$$

and

\begin{equation}
p_{eff} = p - \frac{1}{2} h_{0} ( 1 - 3 \omega )^{\alpha} \rho^{\alpha}
\end{equation}

with $\alpha = \frac{( 1 + 3 \omega )}{2 ( 1 + \omega )}$. Then the above energy conditions can be written as

\begin{eqnarray}
\left. \begin{array}{c}
NEC :~~~ (1 - \omega) \Omega_{m} +(1 + 3 \omega)\geq 0 \\
WEC :~~~ Same as NEC  and \Omega_{m}\geq 0\\
SEC : ~~~3( 1 + \omega )\Omega_{m} - (1 - 3 \omega ) \geq 0 \\
DEC : ~~~Same as WEC
\end{array} \right \}
\end{eqnarray}
with $\Omega_{m} = \rho /3H^{2}$, the density parameter for the matter considered in the Einstein gravity .Now the deceleration parameter $q ( = - ( 1 + \frac{\dot{H}}{H^{2}}))$, is related to $\omega $ and $\Omega_{m}$ by the relation

\begin{equation}
q = (\frac{9 \omega - 1}{4}) + \frac{3 \Omega_{m}}{4}(1 - \omega )
\end{equation}

Thus NEC is satisfied until fluid-1 is in the quintessence era and $\Omega_{m}$ is restricted by the given inequality. Note that the above inequality holds for all $\Omega_{m}$ as long as the normal fluid ( i.e. fluid-1)is not exotic (i.e, satisfies the strong energy condition). The same is true for WEC as well as DEC. However, to satisfy the SEC $\Omega_{m}$ has a lower bound given by $\Omega_{m} \geq \frac{(1 - 3\omega )}{3(1 + \omega )}$. The variation of q has been plotted in figure 1 and the inequalities for NEC and SEC are presented in figures 2 and 3 respectively.
\section{Summary}
The paper deals with recently introduced $f(R,T)$ gravity theory with the restriction of conservation of matter. As a result, although the form of the field  equations remain same but now the test particles move in a geodesics and the choice of the Lagrangian function is not totally arbitrary. We have analyzed three possible choices for $f(R,T)$ and examined whether two familiar matter fields namely electromagnetic field and perfect fluid are permissible or not in this modified gravity theory. It is found that electromagnetic field is not allowed in all the cases. For homogeneous and isotropic model of the universe, the explicit field equations are written for the modified gravity theory with $f(R,T) = R + h(T)$ and it is found that the field equations are equivalent to Einstein gravity with a non-interacting  2-fluid system of which one is the usual perfect fluid in the modified theory while the second fluid (i.e, Fluid-2) is also a barotropic  fluid with constant equation of state and will become exotic when the usual fluid (i.e, Fluid-1) is still a normal fluid. For some specific choice of the equation of state parameter of the usual fluid there are possible cosmological solutions of which one corresponds to big rip singularity. The graph (see figure 1) of q for the variation of $\Omega_{m}$ and $\omega$ shows that there is a natural transition from deceleration to acceleration although we have considered normal fluid (non-exotic), i.e, fluid-1 as the matter source in this modified gravity theory. In particular, if we consider only the baryonic matter (with $\Omega_{m} = 0.04$) as the source of matter field then transition from deceleration to acceleration occurs when $\omega < 0.099$. Thus with the normal fluid model in $f(R,T)$ gravity theory, there is a natural transition from deceleration to acceleration as predicted by recent observations.\\
Also we have analyzed the energy conditions for the modified gravity theory in a general way. For the perfect fluid model of section-III we have shown the validity of the energy conditions both analytically as well as graphically.\\

However, it should be noted that although for some simple choice of f(R, T) we have obtained a possible solution for DE but it is natural to identify the correct class of f(R, T) which are compatible to modern observations [26] as it has been done in f(R) gravity. This is termed as cosmography of f(R, T). In the background of flat FRW model cosmography is related to the taylor series expansion of the scalar factor  around the present time $t_0$ and the first six coefficients in the expansion are [27-29]
$H=\frac{\dot{a}}{a}$, $q=-\frac{1}{aH^2}\frac{d^2 a}{dt^2}$, $j= \frac{1}{aH^3}\frac{d^3 a}{dt^3}$, $s= \frac{1}{aH^4}\frac{d^4 a}{dt^4}$, $l= \frac{1}{aH^5}\frac{d^5 a}{dt^6}$ and $m=\frac{1}{aH^6}\frac{d^6 a}{dt^6}$. They are respectively known as the Hubble parameter, the deceleration parameter, the jerk parameter, the snap parameter, the lerk parameter and the m parameter. These parameters are model independent quantities and are termed as cosmographic set. As a future work, one can analyze the cosmography of f(R, T) gravity to identify the appropriate choices of f(R, T).\\

Moreover, it is interesting to consider hybrid gravity theory related to f(R, T) gravity. In this theory both metric and Palatini formalisms are incorporated in the action [30, 31] and the dynamical scalar corresponding to scalar-tensor representation need to be massive so that it does not care about laboratory and solar system tests and can play an active role in cosmology [30]. Similar to f(R) gravity the action may be chosen as $S= \frac{1}{2\kappa}\int d^4 x \sqrt{-g}[R+f(R, T)]+ S_m$, where R is the Palatini curvature obtained from an independent Palatini connection $\hat{\Gamma}^ \alpha _{\mu\nu}$. This issue may also be considered for future.\\

Finally, f(R, T) gravity theory can be motivated at fundamental level, i.e, at small scales and high energies provided one should take care of quantum field theory formulated on a curved space [32, 33]. Since, at scales comparable to the compton wave length, particles, matter should be quantized so one should employ a semi classical description of gravity and equation (7) is modified as\\

$$f_R R_{\mu\nu}-\frac{1}{2}f(R,\langle T \rangle)g_{\mu\nu}+(g_{\mu\nu}\square -\nabla_{\mu}\nabla_{\nu})f_R= 8\pi \langle T_{\mu\nu} \rangle- f_T(\langle T_{\mu\nu} \rangle+\langle \Theta_{\mu\nu} \rangle)$$\\
where in the arguement of $f_R$ and $f_T$, T should be replaced by $\langle T \rangle$. The expectation value of a quantum stress-energy tensor is defined as [32, 33]\\
$$\langle T_{\mu\nu} \rangle= \langle \Psi \vert \hat{T}_{\mu\nu} \vert \Psi \rangle$$\\
where $\vert \Psi \rangle$ is a quantum state describing the early universe and $\hat{T}_{\mu\nu}$ is the quantum operator associated with the classical energy-momentum tensor of the matter field. In general, a quantized matter field is subject to self interactions as well as it interacts with other fields and with the gravitational background and as a result there are infinities from $\langle T_{\mu\nu} \rangle$. So to obtain a renormalizable theory, one has to introduce infinitely many counterterms in the Lagrangian density [32] of gravity. However, one can construct a truncated quantum theory of gravity by expansion in loops. In this context it should be noted that trace anomaly [33] takes a vital role to deal with infinities in regularization procedures. This is an important issue to deal with for future studies.\\ 

\textbf{Acknowledgements:} The work is done during a visit to IUCAA, Pune (India) under associateship programme. The author is thankful to IUCAA for warm hospitality and facilities at the library. The author also acknowledges the DRS programme of UGC, Govt. of India, in the department of Mathematics, Jadavpur University.\\

\section{references}

$[1]$ A. G. Riess et al., \textit {Astron. J.} \textbf{116}, 1009 (1998) ; S. Perlmutter et al.,{\it Astrophys. J.} {\bf 517}, 565 (1999); P. de Bernardis et al., \textit {Nature} \textbf{404}, 955 (2000); S.Perlmutter et al.,\textit {Astrophys. J.} \textbf{598}, 102 (2003).\\

$[2]$ C. Wetterich,\textit{Nucl. Phys. B} \textbf{302}, 668 (1988) ; B.Ratra, J.Peebles\textit{Phys. Rev. D}\textbf{37}, 321(1988).\\
$[3]$R.R.Caldwell, \textit{Phys. Letts. B}\textbf{545},23(2002); S. Nojiri, S. D. Odinstov\textit{Phys. Letts. B}\textbf{562}, 147(2003); \textit{Phys. Letts .B} \textbf{565}, 1 (2003).\\ \\
$[4]$T. Chiba, T. Okabe, M. Yamaguchi, \textit{Phys.Rev.D}\textbf{62},023511(2000); C.Armendariz-Picon, M. Mukhanov, P. J. Steinhardt \textit{Phys. Rev. Letts}\textbf{85}, 4438 (2000) \textit{Phys. Rev. D}\textbf{63}, 103510 (2001).\\

$[5]$ A.Sen\textit{J. High Energy Phys.}\textbf{04},048(2002);T. Padmanabhan, T.R.Chaudhury\textit{Phys. Rev.D}\textbf{66}, 081301 (2002).\\

$[6]$ E. Elizalde, S. Nojiri, S. D. Odintsov \textit{Phys. Rev. D} \textbf{70}, 043539 (2004); A.Anisimov, E.Babichev, A.Vikman\textit{J.Cosmol.Astropart.Phys}\textbf{06},006(2005).\\

$[7]$ A. Kamenshchik, U. Moschella, V. Pasquier\textit{Phys.Lett.B}\textbf{511},265(2001); M.C.Bento, O.Bertolami, A.A.Sen\textit{Phys.Rev.D}\textbf{66},043507(2002).\\

$[8]$ S.Nojiri and S.D.Odintsov,arXiv:1011.0544(Phys.Rep.,to be published).\\

$[9]$ S.M.Carroll, V.Duvvuri, M.Trodden and M.S.Turner, \textit{Phys.Rev.D}\textbf{70},043528(2004).\\

$[10]$ S.Nojiri and S.D.Odintsov\textit{Phys. Letts. B}\textbf{657},238(2007); \textit{Phys.Rev.D}\textbf{77},026007(2008); arXiv:1008.4275(Prog.Theor.Phys.Suppl.(to be published)); G.Cognola, E.Elizalde, S.Nojiri, S.D.Odintsov, L.Sebastiani and S.Zerbini\textit{Phys.Rev.D}\textbf{77},046009(2008);\textbf{83},086006(2011).\\

$[11]$ S.Capozziello,V.F.Cardone,and A.Troisi, \textit{J. Cosmol. Astropart. Phys}\textbf{08} 001(2006);\textit{Mon.Not.R.Astron.Soc.}\textbf{375},1423(2007).\\

$[12]$ A.Borowiec, W.Godlowski and M.Szydlowski, \textit{Int.J.Geom.Methods Mod.Phys.}\textbf{4},183(2007).\\

$[13]$ C.F.Martins and P.Salucci\textit{Mon.Not.R.Astron.Soc.}\textbf{381},1103(2007).\\

$[14]$ C.G.Boehmer, T.Harko and F.S.N.Lobo, \textit{Astropart.Phys.}\textbf{29},386(2008).\\

$[15]$ C.G.Boehmer, T.Harko and F.S.N.Lobo\textit{J.Cosmol.Astropart.Phys.}\textbf{03},024(2008).\\

$[16]$ T.P.Sotiriou and V.Faraoni\textit{Rev.Mod.Phys.} \textbf{82},451(2010); F.S.N. Lobo, arxiv:0807.1640; S.Capozziello and V.Faraoni \textbf{Beyond Einstein Gravity} \textit{(Springer, N.Y., 2010)}.\\

$[17]$ T.Harko,F.S.N.Lobo, S.Nojiri and S.D.Odintsov\textit{Phys.Rev.D}\textbf{84},024020(2011).\\

$[18]$ M.J.S.Houndio\textit{Int.J.Mod.Phys.D} (to appear); arXiv:1107.3887.\\

$[19]$ M.Sharif and M.Zubair,arXiv:1204.0848(gr-qc).\\

$[20]$ T.Azizi,arXiv:1205.6957(gr-qc).\\

$[21]$ L.D.Landau and E.M.Lifshitz,The Classical Theory of Fields (Butterworth-Heinemann,Oxford,1998).\\

$[22]$ T.Koivisto\textit{Classical and Quantum Gravity}\textbf{23},4289(2006).\\

$[23]$ S.W.Hawing and G.F.R.Ellis,The Large Scale Structure of Space-time (Camb.Univ.Press,Cambridge,England,1973).\\

$[24]$ N.M.Garc\'{i}a , T.Harko,F.S.N.Lobo and J.P.Mimoso \textit{Phys.Rev.D}\textbf{83},104032(2011).\\

$[25]$ M.Visser," Lorentzian Wormholes : From Einstein to Hawking" (Springer,1995).\\

$[26]$ C. M. Will, \textit {Liv. Rev. Relt} \textbf{9}, 3 (2006).\\

$[27]$ A. Aviles, A. Bravetti, S. Capozziello and O. Luongo, \textit{Phys. Rev. D} \textbf{87}, 044012 (2013).\\

$[28]$ S. Weinberg, Cosmology (Oxford University Press, N.Y.2008).\\

$[29]$ U. Alam, V. Sahni, T. D. Saini and A. A. Starobinsky, \textit{Mon. Not. R. Astron .Soc} \textbf{344}, 1057 (2003).\\

$[30]$T. Harko, T. S. Koivisto, F. N. S. Lobo and G. J. Olmo,\textit{Phys. Rev. D} \textbf{85}, 084016 (2012).\\

$[31]$ S. Capozziello, T. Harko, T. S. Koivisto, F. N. S. Lobo and G. J. Olmo \textit{J. Cosmol. Astropart. Phys} \textbf{04}, 011 (2013).\\

$[32]$ S. Capozziello and M. D. Laurentis, \textit{Phys. Rept} \textbf{509}, 167 (2011).\\

$[33]$ N. D. Birrel and P. C. W. Davies, Quantum Fields in Curved Spacetime (Camb. Univ. Press, Camb. U. K. 1982).\\

$[34]$ M. Alves and J. B. Neto, \textit{Braz. J. Phys} \textbf{34}, 531 (2004).

\end{document}